\documentstyle[jkas_eama6]{article}

\runningauthor{TANIGUCHI et al.}
\runningtitle{COSMOS}
\beginpage{55}
\endpage{58}

\def\cm3{~{\rm cm^{-3}}}

\def\lsim{\mathrel{  
        \raise0.3ex\hbox{$<$}\kern-0.75em{\lower0.65ex\hbox{$\sim$}}}}
\def\gsim{\mathrel{
        \raise0.3ex\hbox{$>$}\kern-0.75em{\lower0.65ex\hbox{$\sim$}}}}

\begin{document}

\title{THE HST COSMOS PROJECT: CONTRIBUTION FROM THE SUBARU TELESCOPE}

\author{Yoshiaki Taniguchi$^1$, N. Z. Scoville$^{2, 3}$, D. B. Sanders$^3$,
        B. Mobasher$^4$, H. Aussel$^5$, P. Capak$^2$, M. Ajiki$^1$, 
        T. Murayama$^1$, S. Miyazaki$^6$, Y. Komiyama$^7$, Y. Shioya$^1$,
        T. Nagao$^{1, 8}$, S. Sasaki$^1$, 
        R. Sumiya$^1$, J. Koda$^2$, L. Heinlein$^2$, Y. Hatakeyama$^1$, 
        H. Karoji$^6$, and the COSMOS Team$^9$}

\address{$^1$ Astronomical Institute, Graduate School of Science, Tohoku University,
Aramaki, Aoba, Sendai 980-8578, Japan \\
{\it E-mail: tani@terra.astr.tohoku.ac.jp}}
\address{$^2$ Department of Astronomy, MS 105-24, California Institute of Technology,
 1200 East California Boulevard, Pasadena, CA 91125, USA \\
{\it E-mail: nzs@astro.caltech.edu}}
\address{$^3$ Institute for Astronomy, University of Hawaii, 2680 Woodlawn Drive,
Honolulu, HI 96822, USA \\
{\it E-mail: sanders@galileo.hawaii.edu}}
\address{$^4$ Space Telescope Science Institute, 3700 San Martin Drive,
 Baltimore, MD 21218, USA}
\address{$^5$ Service d'Astrophysique, DSM/DAPNIA/CEA-Saclay, F-91191 Gif-sur-Yvette
  Cedex, France}
\address{$^6$ Subaru Telescope, 650 National Astronomical Observatory of Japan, 
   650 N. A'ohoku Place, Hilo, HI 96720, USA}
\address{$^7$ National Astronomical Observatory of Japan, 2-21-1 Osawa,
   Mitaka, Tokyo 181-8588, Japan}
\address{$^8$ Osservatorio Astrofisico di Arcetri, Largo Enrico Fermi 5, 50125 
  Firenze, Italy}
\address{$^9$ see http://www.astro.caltech.edu/$\sim$cosmos/}
\address{\normalsize{\it (Received February 13, 2005; Accepted February 13, 2005)}}

\abstract{
The Cosmic Evolution Survey (COSMOS) is a Hubble Space Telescope (HST) treasury project.
The COSMOS aims to
perform a 2 square degree imaging survey of an equatorial field in
$I$(F814W) band, using the Advanced Camera for Surveys (ACS).
Such a wide field survey,
combined with ground-based photometric and spectroscopic data,
is essential to understand the interplay between large scale structure,
evolution and formation of galaxies and dark matter.
In 2004, we have obtained high-quality, broad band images
of the COSMOS field ($B, V, r^\prime, i^\prime,$ and $ z^\prime$)
using Suprime-Cam on the Subaru Telescope, and we have started our new
optical multi-band program, COSMOS-21 in 2005.
Here, we present a brief summary of the current status of the COSMOS
project together with contributions from the Subaru Telescope.
Our future Subaru program, COSMOS-21, is also discussed briefly.
}

\keywords{observational cosmology ---
large scale structure -- 
galaxies: evolution}
\maketitle

\section{INTRODUCTION}

\subsection{Background}

The Cosmic Evolution Survey (COSMOS) is an HST treasury project,
 awarded a total of 640 HST orbits,
 to be carried out in two cycles (320 orbits in cycles 12 and 13 each).
This is the largest amount of HST time ever, allocated to a single project.
COSMOS is a 2 square degree imaging survey of an equatorial field in
$I$(F814W) band, using the Advanced Camera for Surveys (ACS). These
observations provide high resolution imaging to map the morphology
of galaxies as a function of environment (overdensity) and epoch, all the
way from high redshift ($z > 3$) to the nearby ($z < 0.5$) universe
(detecting over 2 million galaxies) and
covering a volume in the high redshift universe similar to the Sloan survey
in the low redshift universe.
It is known that substantial Large Scale Structure (LSS) occurs on
scales up to 100 Mpc (comoving), including voids, filaments, groups and
clusters. Therefore, adequately mapping galaxy evolution over the
full range of environments requires multi-waveband data with high spatial
resolution, covering wide areas.

The COSMOS project is fundamental to virtually all areas of galaxy evolution,
identification of different classes of objects,
large scale structure and dark matter evolution, including:

\noindent [1] the evolution of LSS, galaxies, clusters and CDM on
  scales up to $> 10^{14} M_\odot$ as a function of redshift,

\noindent [2] the formation, assembly and evolution of galaxies and star formation as
  a function of LSS environment, morphology and redshift, and

\noindent [3] detailed study of the nature, morphology and clustering properties
of different populations of galaxies [AGN, extremely red objects, Ly$\alpha$
emitters (LAEs), Lyman Break Galaxies(LBGs), star forming galaxies]
and their evolution with redshift, using enormous statistically complete samples.

The large area covered and the depth of the COSMOS, makes it complementary
to other large HST surveys (UDF, GOODS, GEMS, and HDFs); see Fig. 1.
The HST observations of the COSMOS are rapidly progressing, with a total of
over 600 orbits completed so far.
Over the last year, we successfully developed
and partly completed an extensive multi-waveband follow-up program for
the COSMOS (Table 1).
As shown in Table 1, 
the COSMOS project involves major commitments
by non-HST facilities (XMM, CXO, GALEX, and Spitzer), but central
elements are deep multi-color ({\it BVr'i'z'}) imaging with Subaru
($\sim 2\times10^6$ objects -- $\sim$20 nights) and VLT/VIMOS and Magellan/IMACS spectroscopy
($\sim 5\times10^4$ objects -- 540 hrs).
Suprime-Cam on Subaru is an essential element of
COSMOS, uniquely providing
superb multi-color imaging data, well-matched to the depth of the HST 814$I$ 
band imaging.

\begin{figure}[t]
\centerline{\epsfysize=9cm\epsfbox{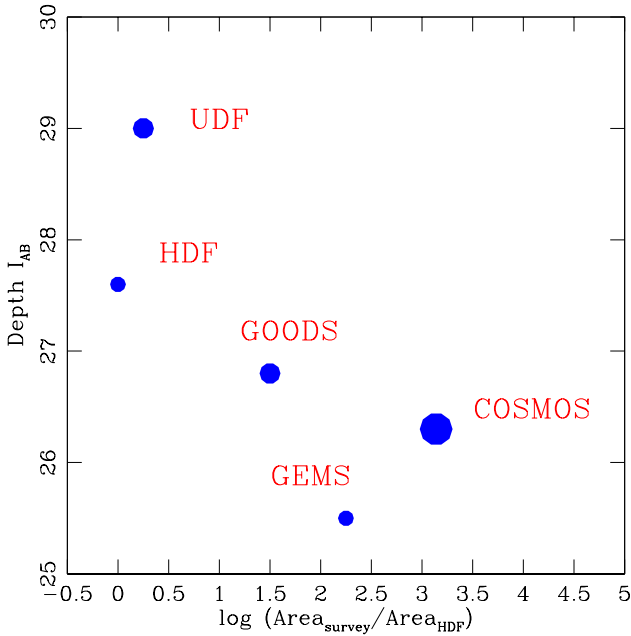}}
\caption{Comparison of the survey depth at $I$ band vs. the survey
area among the deep surveys made with HST; UDF, GOODS, GEMS, HDFs,
and COSMOS.
}
\end{figure}

\subsection{HST Observations}

The HST-ACS observations of the COSMOS field are tailored to 
cover a contiguous area around its center.
We perform one orbit per pointing with 10\% overlap to avoid gaps between
different pointings. Using the ACS pipeline developed
by the Great Observatories Origins Deep Survey (GOODS) team, these data are
immediately reduced and are ready for scientific use.
The HST-ACS observations reach a depth of $I_{\rm AB} = 27.8$
(5$\sigma$) over the entire COSMOS area. We expect to complete all our 
ACS observations by the end of Cycle 13.

While the ACS is used as the primary instrument, we exploit NICMOS ($H$-band/
NIC3) and WFPC2 (UV filter) as parallel instruments. Therefore,
high spatial resolution UV and near-infrared images also exist for parts of
the COSMOS field. These data are also run through the pipeline and reduced.

\begin{center}
\footnotesize
Table 1: {\bf Multi-$\lambda$ COSMOS Data}
\medskip 

~
\begin{tabular}{lcc} \hline
 {Data} & Bands,$\lambda$,Res. & AB mag 5$\sigma$ pt. src \\
\hline
HST-ACS & 814I & 27.8 \\
HST-ACS & 475g & 27.8  \\
HST-ACS & 475g(I) & 27.8  \\
HST-NIC3 & 160W & 22.9(5\% area)  \\
HST-NIC3 & 110W & 23.2(5\% area)  \\
HST-WFPC2 & 450W & 26.4 \\
Subaru-SCam & B, r$^{'}$, z$^{'}$ & 27.4, 27.4, 25.6  \\
Subaru-SCam & NB816 & 25 \\
CFHT-Megacam & u$^{*}$ & 27.4  \\
CFHT-Megacam & u,i$^{*}$ & 26  \\
CFHT-LS & u-z &   \\
NOAO & K$_{s}$ & 22  \\
CFHT & K & 23 ($9^\prime \times 9^\prime$) \\
UH-88 & J & 23.5  \\
GALEX & FUV,NUV & 26.1,25.8  \\
XMM-EPIC & $0.5-10$ keV & 10$^{-15}$ cgs \\
CXO & $0.5-7$ keV & pointed  \\
VLT-VIMOS sp. & (R=200,600) & \#=3000 (I$_{\rm AB}$$<$23) \\
VLT-VIMOS sp. & (R=600) & \#=25000 (I$_{\rm AB}$$<$22.5) \\
VLT-VIMOS sp. & (R=200) & \#=12500 \\ 
              &         & (B$_{\rm AB}$ $<$25.5, 1.4$\leq$z$<$2.5) \\           
Mag.-IMAX sp. & (R=3000) & \#=2000  \\
Keck/GEMINI sp.  & (R=5,000) & \#=4000 (I$<$24) \\
Spitzer-MIPS & 160,70,24$\mu$m & 30,6,0.08 mJy(5$\sigma$) \\
Spitzer-IRAC & 8,6,4.5,3$\mu$m & 12.7,10.3,1.5,0.8 $\mu$Jy(5$\sigma$)  \\
IRAM-30m & 1.2 mm & 1 mJy ($20^\prime \times 20^\prime$)  \\
CSO-Bolocam & 1.1 mm & 3 mJy  \\
VLA-A & 20cm & 24$\mu$Jy(1$\sigma$) \\
VLA-A/C & 20cm & 8$\mu$Jy(1$\sigma$) \\
SZA(full field) & 9 mm & S-Z to $2\times10^{14} M_\odot$  \\
\hline
\end{tabular}
\end{center}
\vskip -0.1in
\footnotesize
\medskip ~

\normalsize

\subsection{Multi-Wavelength Observations}

We have been involved in an extensive observational campaign to use
ground-based and space borne facilities to complement the HST-ACS dataset by
performing multi-waveband photometric and spectroscopic surveys
of the COSMOS field. The present state of ground-based ancillary observations
for the COSMOS are summarized in Table 1.  All the ground-based data
obtained so far are reduced, data quality tests completed and
the multi-waveband catalogs generated. Detailed description and
up-dated progress on the ancillary observations are summarized on
 the COSMOS web page (http://www.astro.caltech.edu/$\sim$cosmos/).
Early results of the COSMOS project were presented in the COSMOS special
session of the AAS meeting in January 2005 (Scoville et al. 2004;
Koekemoer et al. 2004; Lehmann et al. 2004; Mobasher et al. 2004; 
Rhodes et al. 2004; Shopbell et al. 2004; Schinnerer et al. 2004).

\section {CONTRIBUTIONS FROM THE SUBARU TELESCOPE}

\subsection{Five-color, Broad-band Imaging}

The Subaru optical ($BVr'i'z'$) observations, obtained through 8
clear nights in January/February 2004, are completely reduced and
cataloged. These images, taken in an average seeing of 0.7 arcsec,
provide multi-waveband optical data for over $2\times10^6$ galaxies
to $i'=26.5$ mag (5$\sigma$).
Although the HST-ACS observations of the COSMOS field are not yet completed,
using the currently available multi-waveband COSMOS data, we have started
to address the scientific questions which led to the formulation
of the COSMOS project. In the following, we present a brief progress report.

{\it Photometric Redshifts}:
Using the Supreme-Cam optical photometric data, 
we estimated photometric redshifts
for $\sim 10^6$ galaxies detected in the COSMOS field. We also provided the spectral types for all the
galaxies in the Subaru catalog. The photometric redshifts and spectral
types are used to select galaxies for follow-up spectroscopic observations
and to identify clusters.

{\it Extremely Red Objects}:
Combining the near-infrared and optical data, we identified the Extremely Red
Objects (EROs) in the COSMOS field. Study of the HST/ACS morphology of
the EROs shows that they form a heterogeneous population, consisting of
both bulge and disk dominated systems. Moreover, their photometric
redshift distribution shows that the majority of the EROs are located
around $z \sim 1$. Study of the clustering and luminosity function of the EROs
is currently in progress.

{\it Gravitational Lensing}:
For the first time, a comparison is performed between the gravitational
lens candidates on the $I814$-band HST/ACS and $i'$-band ground-based
Subaru images
(see bottom panel of Fig. 2; Miyazaki et al. 2005, in preparation).
We are investigating the efficiency of finding gravitationally lensed systems
with HST, as compared to ground-based Subaru data taken under 0.4 arcsec
seeing conditions.

{\it r' and i'-band drop-outs}:
Samples of r' and i'-band drop-out candidates are identified from the
COSMOS Subaru data. Also, combining the COSMOS $z'$-band and deep $J$-band
data over the $9 \times 9$ arcmin central area of the COSMOS, we have identified
a sample of $z'$-band drop-out galaxies, expected to be at $z\sim 8$.
We are planning spectroscopic observations to confirm the redshift of these
galaxies. This provides an unprecedented sample for statistical studies
of galaxies at $4 < z < 8$.

{\it A cluster of galaxies at z=0.7}:
 Optical identification of
X-ray sources found in our XMM-Newton survey has led to the
discovery of a cluster of galaxies at $z \sim 0.7$ (Fig. 2). 
Follow-up optical spectroscopy has confirmed this cluster.

\begin{figure}[hhh]
\centerline{\epsfysize=10.5cm\epsfbox{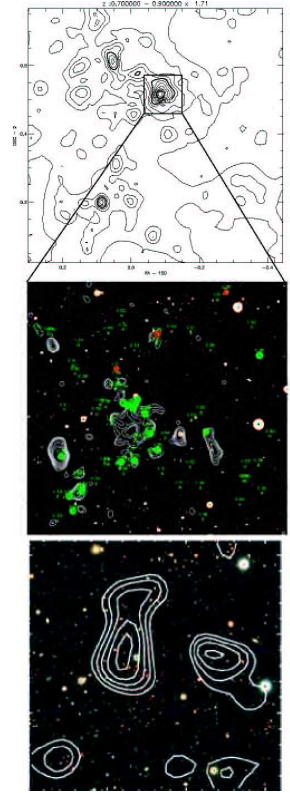}}
\caption{[top] A cluster of galaxies at $z \sim 0.7$ found from our
photo-$z$ catalog around one of XMM sources;  X-ray contours are overlaid on
Suprime-Cam $i'$ band image.
This cluster is spectroscopically confirmed  as a
real cluster by VLT/FORS1 spectroscopy. Large green circles are galaxies
around $z = 0.73$. FOV is 6.7$\times$6.7 arcmin. [middle]
This cluster is also clearly detected in
our weak lensing analysis by Satoshi Miyazaki [bottom].
The contour shows a weak lensing S/N map reconstructed from Suprime-Cam
$i'$-band image whose smoothing scale ($\theta_{\rm G}$) is 30 arcsec.
Here 1500 faint galaxies with $23 < i' < 25.5$ are used for the
reconstruction. The contour starts from S/N = 1 with the interval of 0.5.
FOV is the same as that of left panel.
}
\end{figure}

\subsection{COSMOS-21}

The COSMOS-21 program  aims to extend and complete the ground-based
optical observations of the COSMOS field with deep intermediate and
narrow-band optical imaging of the entire 2~sq.~deg field.
The filters used in the COSMOS-21 project are summarized in Table 2.
As for the intermediate-band filters, see Taniguchi (2004), 
Fujita et al. (2003), Ajiki et al. (2004), Shioya et al. (2005), and 
Yamada et al. (2005).

This COSMOS-21 project 
specifically addresses two major goals:

\noindent [1] The first major COSMIC survey of emission line
(H$\alpha$, Ly$\alpha$ \& [OII]) galaxies with available HST/ACS morphologies,
covering different local environments and sampling the critical range of
redshift for galaxy formation and activity. This allows a wide range of
studies from properties of local ($z < 1$) star forming objects to
proto-galaxies at the re-ionization epoch ($z\sim 7$).

\noindent [2] Major improvement in the accuracy of photometric redshifts (typically
$\delta z/(1+z) < 0.1$) for over a million galaxies at high redshift
($z\sim 7$) down to $i_{\rm AB} \sim 25.5$ (see Figure 3). The value of
intermediate and narrow-band data in improving photometric redshifts is
demonstrated by COMBO-17 survey (Wolf et al. 2003).

The result will be a survey with spectral coverage similar to COMBO-17
but 4 magnitudes deeper, over a larger area and a much more extensive
ancillary (photometric/spectroscopic) dataset. The COSMOS will be a legacy
survey for the world astronomical community for many years to come. With
the possible shortened lifetime of HST, this may be the last and largest survey with
the highest quality imaging for the next 2 decades.

\begin{center}
\footnotesize
Table 2: {\bf A list of filters used in our project}

\medskip ~
\begin{tabular}{cccccc} \hline
 {Filter} & $\lambda_{\rm center}$ (nm) & $\Delta\lambda$ (nm) & $z_{\rm Ly\alpha}$ &
  $z_{\rm [OII]}$ & $z_{\rm H\alpha}$ \\
\hline
NB921 & 920 & 13 & 6.6 & 1.4 & 0.41 \\
NB816 & 815 & 12 & 5.7 & 1.2 & 0.24 \\
IA797 & 797 & 34 & 5.6 & 1.1 & 0.21 \\
IA738 & 738 & 34 & 5.1 & 0.98 & 0.12 \\
IA709 & 709 & 34 & 4.8 & 0.90 & 0.08 \\
IA651 & 651 & 33 & 4.4 & 0.75 & \nodata \\
IA624 & 624 & 31 & 4.1 & 0.67 & \nodata \\
IA598 & 598 & 30 & 3.9 & 0.60 & \nodata \\
IA574 & 574 & 28 & 3.7 & 0.54 & \nodata \\
IA550 & 550 & 27 & 3.5 & 0.48 & \nodata \\
IA527 & 527 & 26 & 3.3 & 0.41 & \nodata \\
IA505 & 505 & 25 & 3.2 & 0.35 & \nodata \\
IA484 & 484 & 24 & 3.0 & 0.30 & \nodata \\
IA464 & 464 & 23 & 2.8 & 0.24 & \nodata \\
IA445 & 445 & 22 & 2.7 & 0.19 & \nodata \\
IA427 & 427 & 21 & 2.5 & 0.15 & \nodata \\
\hline
\end{tabular}
\end{center}
\vskip -0.1in
\medskip ~

\normalsize

\begin{figure}[t]
\centerline{\epsfysize=9cm\epsfbox{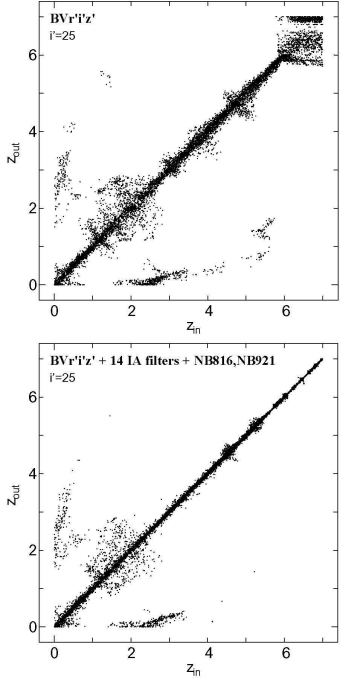}}
\caption{Photo-z simulations. [upper] Case for only broad-band ($BVr'i'z'$)
data; limiting magnitudes are 28 for $B$ and $V$, 27.5 for $r'$, 27.0
for $i'$, and 26.0 for $z'$. [lower] Case for the proposed filter set
($BVr'i'z'$ + 14 IA filters + 2 NB ones: COSMOS-21); limiting magnitudes are
27.5 for $B$, 27.0 for $V$, 27.0 for $r'$, 26.6
for $i'$, 25.6 for $z'$ (our current data obtained with Suprime-Cam),
26.0 for the 14 IA filters and 25.5 for the two NB ones.
}
\end{figure}

\acknowledgements
We would like to thank all the members of the COSMOS team.

\end{document}